\documentclass[%
 aip,
 amsmath,amssymb,
 reprint,%
]{revtex4-1}

\usepackage{graphicx}
\usepackage{dcolumn}
\usepackage{bm}

\usepackage[utf8]{inputenc}
\usepackage[T1]{fontenc}
\usepackage{mathptmx}
\usepackage{etoolbox}
\usepackage{color} 

\makeatletter
\def\@email#1#2{%
 \endgroup
 \patchcmd{\titleblock@produce}
  {\frontmatter@RRAPformat}
  {\frontmatter@RRAPformat{\produce@RRAP{*#1\href{mailto:#2}{#2}}}\frontmatter@RRAPformat}
  {}{}
}%
\makeatother
\begin{document}

\preprint{AIP/123-QED}

\title{Motion-Tracking Terahertz Time-Domain Spectroscopy Enabled by Time-Programmable Frequency Combs for Moving-Object Sensing}
\author{Riku Shibata}
 \affiliation{Department of Physics, Faculty of Science and Technology, Keio University, Yokohama, 223-8522, Japan}
\author{Shun Fujii}
 \affiliation{Department of Physics, Faculty of Science and Technology, Keio University, Yokohama, 223-8522, Japan}
\author{Tomofumi Ikari}
 \affiliation{College of Engineering, Nihon University, Koriyama 963-8642, Japan}
\author{Shinichi Watanabe}%
 \email{watanabe@phys.keio.ac.jp}
\affiliation{Department of Physics, Faculty of Science and Technology, Keio University, Yokohama, 223-8522, Japan}%

\date{\today}

\begin{abstract}
Terahertz time-domain spectroscopy (THz-TDS) provides direct access to both spectral information and time-of-flight features, making it attractive for dynamic sensing. However, conventional high-speed THz-TDS methods typically rely on a fixed delay trajectory, limiting their ability to selectively acquire only the relevant temporal window while tracking target motion. Here, we demonstrate motion-tracking THz-TDS based on time-programmable frequency combs (TPFCs), in which the THz measurement window is actively locked to a moving time-domain waveform. By continuously modulating the phase-lock set point of one TPFC, we perform real-time apodized acquisition of a specific temporal window centered on the main THz peak at rates up to 308 Hz. Simultaneously, the residual peak position within the window is detected in real time and fed back to the other comb, enabling the measurement window to follow target displacements exceeding the original acquisition window. The feedback command together with the residual peak position further enables reconstruction of the relative target displacement. As a proof-of-concept demonstration, we use a moving gold mirror to simulate respiration- and heartbeat-related motion superimposed on a much larger displacement representing body motion, and successfully resolve the two vital-sign-related components from the measured displacement. This work extends arbitrary-delay control with TPFCs to THz-TDS and establishes a motion-compensated spectroscopy platform for simultaneous displacement tracking and THz waveform acquisition.
\end{abstract}

\maketitle

\section{\label{sec:level1}Introduction}

Terahertz (THz) sensing has emerged as a high-resolution sensing modality at the interface between microwave and optical technologies, enabling noncontact inspection, spectroscopic identification, and material characterization in a wide range of applications \cite{Tonouchi:2007}. Compared with microwave and millimeter-wave radiation, the shorter wavelengths of THz waves provide higher directivity and improved spatial localization \cite{Huang:2023}. At the same time, THz waves can penetrate clothing \cite{Bjarnason:2004}, and their non-ionizing nature and high spatial localization make them suitable for noncontact human sensing, including security screening \cite{Sheen:2001,Appleby:2007, Ikari:2025} and physiological monitoring \cite{Petkie:2008}. Their ability to penetrate dielectric materials, such as plastics \cite{Hu:1995} and paints \cite{Yasui:2005}, further enables the inspection of concealed objects, packaged materials, and coating thicknesses, highlighting their utility for industrial nondestructive evaluation.

Among various THz sensing modalities, THz time-domain spectroscopy (THz-TDS) uses ultrashort laser pulses and optical delay scanning to directly measure the time-domain electric-field waveform of THz pulses, providing unique advantages in both time-of-flight analysis and spectroscopy \cite{Koch:2023}. This time-domain measurement enables time-of-flight-based tomography with high signal-to-noise ratios by resolving reflected echoes from different interfaces \cite{Mittleman:1997}. The resulting temporal separation enables isolation of the target signal from undesired delayed components, reducing artifacts from overlapping echoes that often complicate continuous-wave-based microwave and terahertz sensing. However, most THz-TDS measurements implicitly assume that the target position remains fixed during data acquisition. This assumption becomes particularly problematic in reflection geometry when the target object is not stationary but in motion, because target motion shifts the reflected THz waveform in delay and causes the measurement window to slip relative to the sample. Such situations arise in many practical sensing scenarios: security screening \cite{Ikari:2025} and physiological sensing \cite{Oyamada:2021,Wang:2025} often involve unavoidable motion of the human body, whereas industrial inspection frequently requires measurements of target materials transported on conveyor systems \cite{Ellrich:2020}, where the height or sensor-to-sample distance of the object can vary continuously during measurement. To fully exploit THz-TDS in practical reflection measurements, it is therefore essential to track target motion and keep the desired temporal region of the reflected waveform within the measurement window.

To realize such motion-tracking THz-TDS, high-speed acquisition of THz-TDS waveforms is a primary requirement. Various high-speed THz-TDS techniques have been developed to accelerate optical delay scanning and increase waveform acquisition rates \cite{Li2023}, including rapidly scanned mechanical delay lines \cite{Guerboukha2015,Vieweg2014}, acousto-optic delay lines \cite{Urbanek2016}, optical sampling by laser cavity tuning \cite{Wilk2011,Molteni2023}, and single-laser polarization-controlled optical sampling \cite{Kolano2018}. Dual-laser approaches, such as asynchronous optical sampling (ASOPS) \cite{Janke2005,Yasui2005,Yasui2012,Good2015,Kliebisch2016,Okano2022} and electronically controlled optical sampling \cite{Kim2010,Dietz2014}, eliminate mechanical delay scanning and enable faster waveform acquisition. However, existing methods do not allow arbitrary shifting of the temporal origin of the THz-TDS waveform, and thus cannot keep the measurement window focused on the desired waveform region while continuously following target motion.

In this work, we report motion-tracking THz-TDS, which selectively acquires arbitrary temporal regions of the reflected THz-TDS waveform while following the motion of a target object. Our method uses a pair of time-programmable frequency combs (TPFCs) \cite{Caldwell:2022}, whose pulse timing and carrier phase can be digitally controlled without a fixed trigger-defined temporal origin ($t = 0$), allowing the temporal origin to be shifted freely. Although TPFCs have been used for ranging \cite{Caldwell:2022} and programmable near-infrared dual-comb spectroscopy with flexible temporal sampling \cite{Giorgetta2024,Giorgetta2024-2}, their application to THz-TDS has not yet been reported. Moreover, real-time apodized acquisition of time-domain waveforms while actively tracking target motion remains unexplored. The proposed system keeps the desired THz waveform region within a narrow measurement window during target motion, enabling high-speed acquisition of reflected THz waveforms and simultaneous extraction of displacement information from the control signal. We demonstrate centimeter-scale tracking of a moving mirror and the detection of simulated vital-sign-related motion superimposed on large displacements mimicking body motion. These results establish a THz-TDS approach that enables simultaneous acquisition of THz waveform and spectral information together with displacement information in a reference frame that follows the target motion.

\section{\label{sec:level2}Experimental setup}

\begin{figure*}
\includegraphics[width=\linewidth]{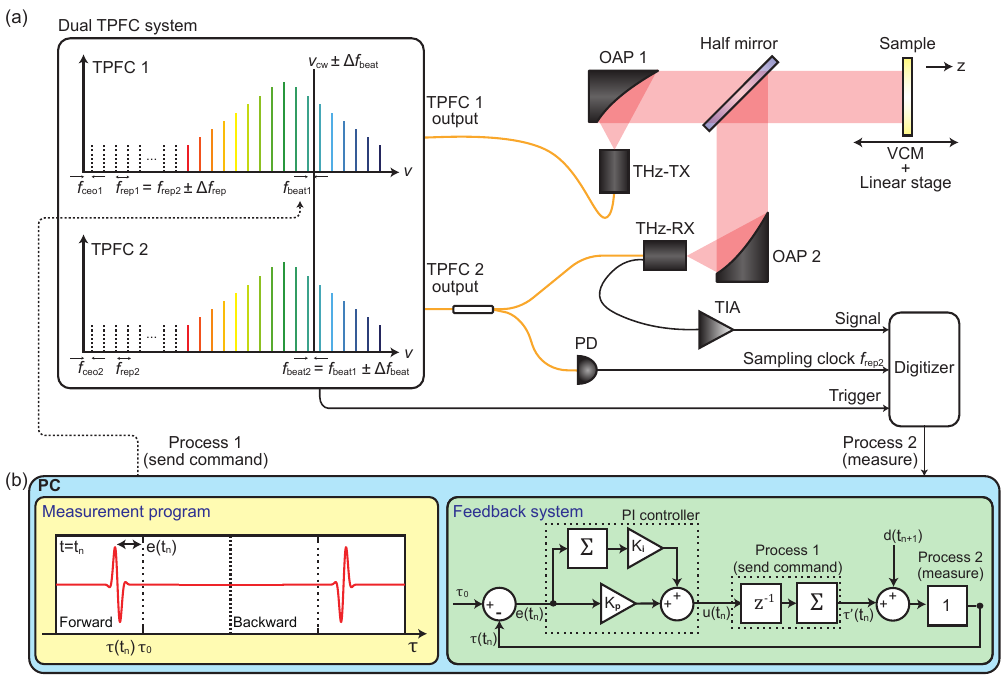}
\caption{\label{fig1} Overview of motion tracking THz time domain spectroscopy system. (a) Schematic illustration of the optical spectra of the dual TPFC system and experimental setup. The two TPFCs are tightly phase-locked to each other via a continuous-wave external-cavity laser with an optical frequency of $\nu_{\mathrm{cw}}$. Orange lines indicate the optical fibers and black solid lines indicate the electric cables. THz-TX: THz transmitter, THz-RX: THz receiver, OAP: Off-axis parabolic mirror, VCM: Voice coil motor, TIA: Transimpedance amplifier, PD: Photodetector. (b) Schematic of motion-tracking THz-TDS implemented on the PC. The yellow box shows one THz waveform measured by real-time apodized acquisition. The green box shows the feedback system, consisting of a PI controller, Process~1 for controlling $f_{\mathrm{beat1}}$ with a one-frame delay, and Process~2 for measuring the THz peak position. Sample-motion-induced peak shifts are treated as the disturbance $d(t)$.}
\end{figure*}

Figure 1 shows a schematic of the motion-tracking THz-TDS system. The system employs two tightly phase-locked TPFCs for THz pulse generation and detection, as illustrated in Fig.~\ref{fig1}(a). The frequency-locking scheme of the two TPFCs was identical to that used in our previous architecture \cite{Okano2022}, except for the time-programmed control, which represents a key advance of the present work. The output of TPFC~1 was fiber-coupled to a THz transmitter (THz-TX; Fraunhofer HHI) to generate THz pulses. The generated THz pulses were collimated by an off-axis parabolic mirror (OAP), transmitted through a 4-inch silicon wafer with a thickness of 525-$\mu$m used as a half mirror, and directed onto the sample. The THz pulses reflected from the sample were subsequently reflected by the half mirror and focused by another OAP onto a THz receiver (THz-RX; EK-000782, TOPTICA Photonics AG). The THz-RX, gated by TPFC~2, detected the THz electric field as a photocurrent signal. The photocurrent was amplified using a transimpedance amplifier (SA-605F2, NF Corp.) and digitized at the repetition frequency of TPFC~2, $f_{\mathrm{rep2}}$, using a digitizer (M2p.5962-x4, Spectrum Instrumentation GmbH) installed in the control PC.
The repetition frequency $f_{\mathrm{rep2}}$ was set to 61.7 MHz throughout this study.

\subsection{\label{sec:level2}Real-time Apodized Terahertz Time-Domain Waveform Detection Using Time-Programmable Frequency Combs}

Apodized terahertz time-domain waveform detection was implemented using time-programmed control. Repetitive modulation and feedback control of the optical delay between the two TPFCs were achieved by dynamically controlling the phase-locking conditions of their optical beat frequencies, $f_{\mathrm{beat1}}$ and $f_{\mathrm{beat2}}$, with respect to a continuous-wave external-cavity laser. First, 
these optical beat frequencies, together with the carrier-envelope-offset frequencies $f_{\mathrm{ceo1}}$ and $f_{\mathrm{ceo2}}$, were frequency-locked such that $f_{\mathrm{rep1}} = f_{\mathrm{rep2}}$, where $f_{\mathrm{rep1}}$ denotes the repetition frequency of TPFC~1. Under this condition, the relative timing between the two TPFC pulse trains remained fixed without temporal scanning, thereby defining a provisional temporal origin. 
Next, the optical delay between the two TPFCs was scanned once over a wide temporal range by varying $f_{\mathrm{beat2}}$ while keeping $f_{\mathrm{beat1}}$, $f_{\mathrm{ceo1}}$, and $f_{\mathrm{ceo2}}$ constant to find the THz peak. Subsequently, $f_{\mathrm{beat1}}$ was tuned so that the identified THz peak coincided with the updated temporal origin, and was then held fixed.
Then, THz-TDS measurements were performed within a fixed sampling window around the updated temporal origin by periodically varying $f_{\mathrm{beat2}}$ over a range of $\pm 3.61$ MHz. This modulation caused $\Delta f_{\mathrm{rep}} \equiv f_{\mathrm{rep2}}-f_{\mathrm{rep1}}$ to alternate between $+1.16$ and $-1.16$ Hz, enabling repeated acquisition of real-time temporally apodized waveforms around the updated temporal origin. We define the slew rate of the system as the rate of change of the optical delay between the two lasers, converted to the THz time axis, which is $\pm$~18.8 ns/s.

This architecture takes advantage of the free-form operation introduced in Ref.~\cite{Giorgetta2024}, which enables real-time apodization by tailoring the temporal scanning trajectory. In the present THz-TDS implementation, the delay scan can be confined to a narrow temporal region around the THz pulse, thereby enabling the relevant THz waveform to be acquired within a substantially shorter acquisition time.
In this work, the $f_{\mathrm{beat2}}$ modulation frequency was set to 15.4 and 154 Hz, corresponding to THz-TDS temporal windows of 610 and 61 ps, respectively. The corresponding scan times for one complete forward-and-backward THz-TDS waveform scan were 65 and 6.5 ms, respectively. This capability provides a significant advantage over the previous ASOPS-based THz-TDS system \cite{Yasui2005, Okano2022}, in which the entire temporal window of $f_{\mathrm{rep1}}^{-1}=16.2$ ns had to be scanned, requiring $\Delta f_{\mathrm{rep}}^{-1}=0.86$ s to acquire a single THz-TDS waveform.

\subsection{\label{sec:level2}Motion-tracking THz-TDS}
Next, we introduce motion-tracking real-time apodization of THz-TDS, which is a key element of the present work. Figure~\ref{fig1}(b) illustrates the feedback procedure. The yellow box schematically shows the terahertz waveform acquired by the measurement during one forward- and backward-scan cycle under real-time apodization.
In the present experiment, the $f_{\mathrm{beat2}}$ modulation frequency was set to 154 Hz, corresponding to a THz-TDS temporal window of 61 ps and $\Delta t=6.5$~ms for acquiring one pair of forward- and backward-scan THz-TDS waveforms.
We set the feedback setpoint $\tau_0$ at the center of the temporal window for the forward-scanned waveform.
For each acquisition of a real-time apodized waveform, the peak position of the THz waveform in the forward-scanned waveform, $\tau(t)$, is extracted from the measured waveform, and the error signal,
\begin{equation}
e(t) = \tau_0 - \tau(t).
\label{eq1}
\end{equation}
is calculated.
In motion-tracking THz-TDS, feedback control of $f_{\mathrm{beat1}}$ is used to adjust the relative timing between the two lasers, thereby redefining the temporal origin so as to maintain $e(t)=0$. In other words, the peak position $\tau(t)$ of the forward-scan waveform is kept at a prescribed set point $\tau_0$ within the temporal window.

The green box shows a block diagram of the discrete-time feedback control system. The feedback control was implemented in discrete time with an interval of $\Delta t$, such that $t=t_n=n\Delta t$, where $n$ is an integer. The measured peak position $\tau(t_n)$ is compared with the set point $\tau_0$ to obtain the error signal $e(t_n)$.
The error signal is processed by a PI controller to generate the control signal $u(t_n)$, according to
\begin{equation}
u(t_n) = K_p e(t_n) + K_i\Delta t \sum_{j=0}^n e(t_j),
\label{eq2}
\end{equation}
where $K_p$ and $K_i$ are the proportional and integral gains, respectively.
We set $K_p=15.4~\mathrm{s}^{-1}$ and $K_i=119~\mathrm{s}^{-2}$ based on step-response measurements to achieve a favorable balance between settling performance and response speed. Further details of the gain-tuning procedure are provided in Appendix A.

Process~1 in the feedback loop (green box in Fig.~\ref{fig1}(b)) adjusts the locking phase of $f_{\mathrm{beat1}}$ according to the calculated control signal $u(t_n)$. This adjustment shifts the time origin, and hence the peak position of the THz waveform within the temporal window, by $u(t_n)\Delta t$. 
This adjustment is completed within one frame interval, $\Delta t$, as long as the THz peak lies within the measured temporal window, because $f_{\mathrm{beat1}}$ is varied at the same absolute slew rate as $f_{\mathrm{beat2}}$.
In practice, we found that the feedback command issued at $t=t_n$ takes effect at $t=t_{n+1}$, resulting in a one-frame delay, as indicated by $z^{-1}$ in Fig.~\ref{fig1} (b). Accordingly, instead of integrating $u$ up to the current time $t=t_n$, only the control signals up to the preceding frame, $t=t_{n-1}$, contribute to the updated peak position at $t=t_n$, which is given by
\begin{equation}
\tau'(t_n) = \Delta t\sum_{j=0}^{n-1} u(t_j).
\label{eq3}
\end{equation}
Meanwhile, unintended motion of the sample introduces a temporal displacement denoted by $d(t_{n+1})$. Process~2 in the feedback loop then measures the resulting peak position of the THz waveform, given by
\begin{equation}
\tau(t_{n+1}) = \tau'(t_n) + d(t_{n+1}).
\label{eq4}
\end{equation}
The measured peak position $\tau(t_{n+1})$ serves as the input to the next feedback-control cycle.
This discrete-time feedback control maintains the THz peak position at the prescribed temporal position, thereby enabling practical real-time motion tracking. The control signal $u(t_n)$ used for feedback and the corresponding error signal $e(t_n)$ were recorded throughout the measurement.
The frequency response of the present motion-tracking control system is characterized in Appendix B.

\section{\label{sec:level2} Results and discussions}
\subsection{\label{sec:level2} Real-time apodized THz-TDS}
\begin{figure*}
\includegraphics[width=0.67\linewidth]{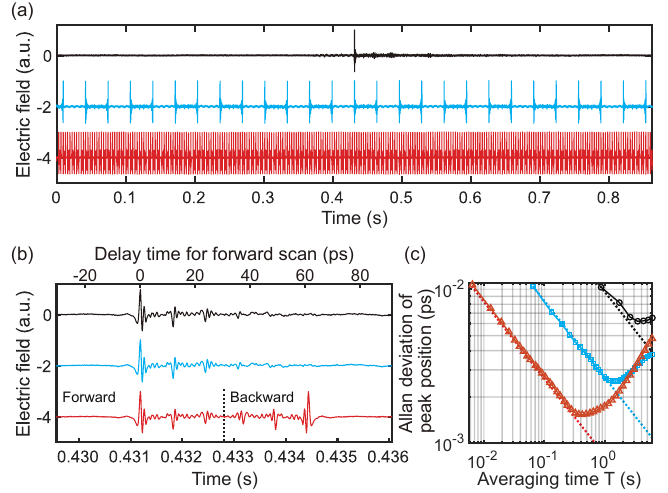}
\caption{\label{fig2} (a) Measured THz waveforms. From top to bottom: the waveform obtained using ASOPS (black), real-time apodization with a 610-ps time window (blue), and real-time apodization with a 61-ps time window (red). All waveforms were smoothed using a 1000-point moving average.
(b) Enlarged view of the THz waveforms in (a) around the time origin. The lower and upper horizontal axes indicate measurement time and delay time, respectively. For the red waveform, the delay time increases during the forward scan and decreases after the turning point marked by the black dashed line; thus, the upper axis applies directly only to the forward-scan portion. (c) Allan deviation of the THz peak position along the delay-time axis for the forward-scan waveforms. Dashed lines indicate the expected $T^{-1/2}/\sqrt{N}$ scaling for $N=1$, 13, and 133, corresponding to the numbers of forward-scan waveforms acquired within one conventional ASOPS acquisition period.}
\end{figure*}
Figure 2(a) shows the temporal waveforms of the THz electric field acquired using conventional ASOPS and the real-time apodized methods with a temporal window of 610 ps and 61 ps. 
For conventional ASOPS, $\Delta f_{\mathrm{rep}}$ was set to 1.16 Hz, approximately equal to the magnitude $|\Delta f_{\mathrm{rep}}|$ used under the real-time-apodized condition. 
In conventional ASOPS, only one complete waveform was acquired within a measurement time of approximately 0.86 s, covering the full 16.2-ns temporal window along the delay-time axis determined by $1/{f_{\mathrm{rep1}}}$. By contrast, real-time apodization enabled repeated acquisition of the temporal region of interest, yielding as many as 27 and 266 waveforms within the same measurement time for temporal windows of 610 and 61 ps, respectively. Thus, by restricting the measurement to the required temporal window, real-time apodization substantially increased the number of waveforms acquired per unit time.

Figure 2(b) shows an enlarged view of the detected THz electric-field waveforms around the temporal origin. 
The displayed measurement-time interval corresponds to a delay-time span of 122 ps for all three waveforms. For the red waveform obtained with the 61-ps apodization window, this interval contains one complete bidirectional scan cycle. 
The backward-scan waveform is the time-reversed counterpart of the forward-scan waveform and can be converted to the same delay-time orientation by reversing its delay-time axis. Notably, almost no degradation of the red waveform was observed, even near the turning point. This result indicates that fluctuations in $f_{\mathrm{rep1}}$ during the switching of $f_{\mathrm{beat2}}$ were small.

In the apodized measurement, the waveform was not only free from distortion near the turning point, but its THz peak position also remained stable from one acquisition to the next. To quantify this stability, we evaluated the Allan deviation of the THz peak position. For each forward scan, the peak position along the delay-time axis was extracted over the entire measurement window for the conventional ASOPS measurement and within the apodized time window for the real-time apodized measurement. The measurements were continuously repeated for 60 s, and the Allan deviation was calculated using the MATLAB allanvar function. 
Figure 2(c) shows the Allan deviation of the peak position as a function of averaging time under the three measurement conditions. At the shortest averaging time, corresponding to the acquisition time of a single waveform, the Allan deviation is approximately 10 fs for all three conditions. This result demonstrates that real-time apodization preserves essentially the same short-term peak-position stability as conventional ASOPS, even though the apodized time window is generated by rapidly switching $f_{\mathrm{beat2}}$. 
The Allan deviation decreases approximately as $T^{-1/2}$ with increasing averaging time $T$ up to a certain timescale. Because the three measurements exhibit nearly identical initial Allan deviations, the improved precision is governed primarily by the number of accumulated waveforms. Indeed, the measured Allan-deviation curves approximately follow the expected $T^{-1/2}/\sqrt{N}$ dependence, where $N=1$, 13, and 133 denote the numbers of forward-scan waveforms acquired within $1/\Delta f_{\mathrm{rep}}$, for the three measurements, respectively. This result demonstrates the enhanced measurement precision enabled by the apodized THz-TDS framework.
Note that the Allan deviation reaches a minimum at a certain averaging time and then increases, presumably because unavoidable long-term fluctuations become significant during prolonged repetitive measurements. This minimum occurs at a shorter averaging time for measurements with a higher switching rate, suggesting that switching-related fluctuations accumulate more rapidly and eventually limit the benefit of further averaging. Nevertheless, the best timing stability of 1.5 fs corresponds to a displacement stability of 225 nm, which is sufficient for detecting small displacements in practical applications.

\subsection{\label{sec:level3}Motion-tracking Real-time apodized THz-TDS}
\begin{figure}
\includegraphics[width=\linewidth]{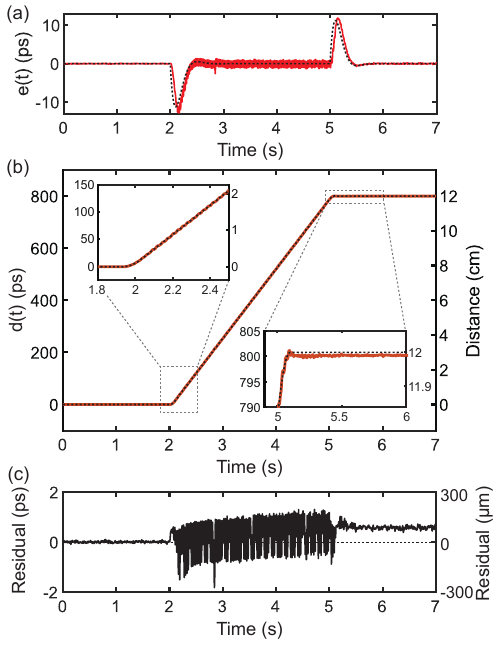}
\caption{\label{fig3}(a) Measured peak-position error $e(t)$ (red curve) and the analytical prediction obtained from the feedback model (black dashed curve). (b) Reconstructed temporal shift $d(t)$ induced by the mirror motion (red curve) and the prescribed temporal-shift trajectory corresponding to the commanded mirror displacement (black dashed curve). The right vertical axis indicates the corresponding mirror displacement. The insets show enlarged views around the onset and termination of the mirror motion. (c) Residual obtained by subtracting the prescribed trajectory from the reconstructed temporal shift in (b). The right vertical axis indicates the corresponding displacement residual.}
\end{figure}

We performed motion-tracking THz-TDS measurements to demonstrate the practical applicability of reflection-mode THz-TDS to moving objects. Motion tracking was initiated after positioning the THz peak position at the center of the forward-scan temporal window. 
The long-range mirror displacement was implemented over a total measurement time of 7 s. The mirror was initially held stationary for 2 s, then displaced by 12 cm at a velocity of 4 cm/s for 3 s along the positive $z$-direction (indicated in Fig.~\ref{fig1}(a)), and finally held stationary again for 2 s. Without motion tracking, the accessible displacement range was limited to $\pm 4.5$~mm by the accessible THz-TDS time window of $\pm 30.5$~ps, and a 12-cm displacement would therefore cause the reflected THz waveform to quickly move outside the measurement window.

During motion tracking, for each pair of forward and backward scans, we determined $\tau(t)$ from the forward scan and calculated the error as $e(t)=\tau_0-\tau(t)$. 
Feedback control was then applied to maintain $e(t)=0$. As shown in Fig.~\ref{fig2}(c), the THz peak position is stabilized with a precision of 10 fs in a single scan, corresponding to a mirror-position precision of 1.5 $\mu$m.
The red curve in Fig.~\ref{fig3}(a) shows the measured peak-position error $e(t)$ during motion-tracking THz-TDS. Although feedback control was applied to maintain $e(t)=0$, two transient peaks appeared near $t=2$~s and $t=5$~s, with amplitudes of approximately $-12$~ps and $+12$~ps, respectively.
These transient errors arise from the transient response of the PI feedback system to the abrupt changes in the mirror velocity at these times. To quantitatively describe these transient errors, we derived an approximate transfer function from the block diagram shown in Fig.~\ref{fig1}(b), neglecting the delay unit, and obtained an analytical expression for $e(t)$ by modeling the mirror motion as a ramp input, as described in Appendix~C. The resulting analytical solution for $e(t)$ is plotted as the black dashed curve in Fig.~\ref{fig3}(a). The measured values of $e(t)$ agree well with the theoretical prediction, confirming that the motion-tracking THz-TDS system responds as expected for the given control parameters. Importantly, despite the approximately 800-ps temporal shift induced by the 12-cm mirror motion, $e(t)$ remained within approximately $\pm 12$~ps throughout the entire measurement period. This result indicates that the THz pulse peak was successfully maintained within the $\pm 30.5$~ps temporal window even while the mirror was moving, thereby demonstrating the effectiveness of the motion-tracking feedback control.

The actual temporal shift in the THz peak position induced by the mirror motion can be reconstructed from Eqs.~\eqref{eq3} and \eqref{eq4} as,
\begin{align}
d(t_{n})&=-\tau'(t_{n-1})
-e(t_{n}) \notag\\
&=-\Delta t\sum_{j=0}^{n-2} u(t_{j})-e(t_{n}).
\label{eq5}
\end{align}
Figure~\ref{fig3}(b) shows the reconstructed $d(t_{n})$ together with the time-shift trajectory corresponding to the prescribed stage motion, including the displacement, acceleration, and velocity. The reconstructed trajectory is in overall agreement with the actual mirror motion.

To examine the small discrepancies in more detail, Fig.~\ref{fig3}(c) shows the difference between the reconstructed $d(t_{n})$ and the prescribed mirror trajectory. Two differences can be observed between the two traces.
First, after the motion was completed, a residual discrepancy of approximately 80~$\mu$m remained. 
To address this problem, we performed conventional ASOPS-based THz-TDS for reflection ranging with much slower detection frequency at 4.7 Hz, and a similar discrepancy of approximately 80~$\mu$m was also observed. This discrepancy is therefore attributed not to the motion-tracking framework, but to angular misalignment between the translation axis of the linear stage and the optical axis of the THz beam. Assuming that the measured displacement corresponds to the projection of the stage displacement onto the THz-beam axis, an 80-$\mu$m discrepancy over a displacement of 12~cm corresponds to an angular misalignment of approximately $2^\circ$, which is reasonable for the present experimental configuration.
Second difference is that it contains oscillatory components with a frequency of approximately 36 Hz during the mirror motion. This oscillation may originate from unintended oscillations in the feedback loop caused by the feedback delay. 
Nevertheless, because this noise is concentrated within a specific frequency range, it can be readily removed by Fourier filtering, as demonstrated in the next section.


Finally, we discuss the tracking range and velocity limits of motion-tracking THz-TDS. Because motion tracking is achieved by continuously varying the locking phase of $f_{\mathrm{beat1}}$, the tracking distance has no intrinsic upper limit. By contrast, the maximum trackable mirror velocity is limited by the transient peak-position error $e(t)$, which must remain within half the width of the temporal window.
In the present experiment, with a temporal window of $\pm 30.5$~ps and the control parameters used here, the maximum trackable mirror velocity was analytically estimated to be $10.8$~cm/s, taking into account the sudden change in velocity. Details of the analytical derivation are provided in Appendix C. This derivation also provides a useful guideline for achieving higher tracking velocities through optimization of the control parameters.

\subsection{\label{sec:level5}Model demonstration of non-contact vital-sign sensing}
\begin{figure*}
\includegraphics[width=\linewidth]{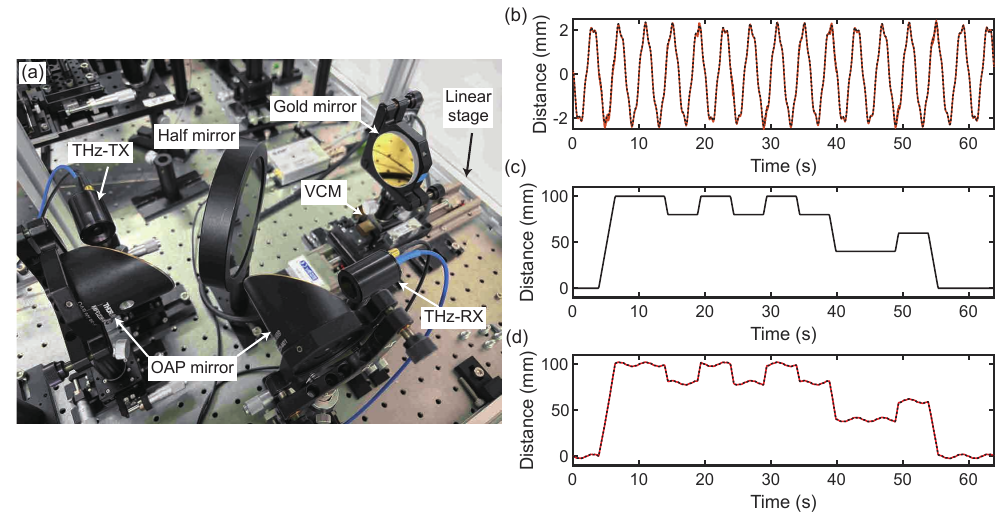}
\caption{\label{fig4} (a) Photograph of the experimental setup for a model demonstration of vital-sign sensing. (b) Experimentally extracted VCM-related displacement waveform obtained by motion-tracking THz-TDS (red curve) and prescribed displacement waveform used to drive the VCM (black dashed curve). (c) Prescribed linear-stage displacement waveform. (d) Reconstructed displacement obtained by motion-tracking THz-TDS (red curve) and reference total displacement calculated from the prescribed VCM displacement and the linear-stage displacement (black dashed curve).}
\end{figure*}

The results presented above demonstrate that motion-tracking THz-TDS combines micrometer-scale displacement resolution, a high waveform acquisition rate, and the ability to continuously follow a moving target in reflection-mode measurements. These characteristics are particularly advantageous for noncontact vital-sign sensing. Chest-wall motion associated with respiration and heartbeat occurs on the millimeter and submillimeter scales\cite{Hao2025}, while unavoidable body motion during measurement can produce substantially larger displacements that may cause the THz waveform to leave the accessible temporal window in conventional THz-TDS. Motion-tracking THz-TDS is therefore well suited to detecting small vital-sign-related displacements in the presence of larger body motion. To demonstrate this potential application, we conducted a model experiment using a gold mirror representing the human chest surface, mounted on a voice coil motor (VCM) driven at frequencies corresponding to respiration and heartbeat.

Figure~\ref{fig4}(a) shows the photograph of the experimental setup. A gold mirror mounted on a VCM was used to reproduce chest-wall displacement caused by respiration and heartbeat. The VCM was driven according to the displacement waveform shown in black dashed curve in Fig.~\ref{fig4}(b). The signal waveform was composed of the superposition of two sinusoidal waves corresponding to respiration and heartbeat:
\begin{equation}
x_\mathrm{VCM}=A_b\sin(2\pi f_b t)+A_h\sin(2\pi f_h t),
\label{eq6}
\end{equation}
where $A_b=2.1 $~mm and $A_h=0.21$~mm are the amplitudes of the respiratory and heartbeat components, respectively, and $f_b=0.25$~Hz and $f_h=1.2$~Hz are their corresponding frequencies. The parameters were selected within the typical ranges of adult chest-wall motion\cite{Hao2025}.
The VCM was mounted on a linear stage to simulate large-scale unintended body motion. 
With the VCM driven according to Eq.~\eqref{eq6}, the linear stage was simultaneously moved over a range of up to 10 cm according to the prescribed displacement waveform shown in Fig.~\ref{fig4}(c).
The black dashed curve in Fig.~\ref{fig4}(d) shows the prescribed total displacement of the gold mirror, resulting from the superposition of the millimeter-scale VCM oscillation and the centimeter-scale motion of the linear stage. The accessible THz-TDS time window without position feedback was $\pm 30.5$~ps, corresponding to a sample displacement range of $\pm 4.5$~mm (the same values as those used in Sec.~III B).
Thus, the motion-tracking THz-TDS measurement was tested under conditions in which the prescribed sample motion exceeded the accessible displacement range without position feedback.

The red curve in Fig.~\ref{fig4}(d) shows the measured displacement $D(t)$ of the mirror during the motion-tracking THz-TDS measurement. The displacement was reconstructed from $d(t)$ in Eq.~\eqref{eq5} using the relation $D(t)=c d(t)/2$, where $c$ is the speed of light in air. 
The reconstructed displacement closely follows the reference trajectory shown by the black dashed curve, demonstrating that the combined VCM and linear-stage motion was successfully recovered. The red curve in Fig.~\ref{fig4}(b) shows the displacement after subtracting the prescribed stage displacement (Fig.~\ref{fig4}(c)) from the reconstructed displacement (Fig.~\ref{fig4}(d)). The extracted waveform agrees well with the prescribed VCM displacement, demonstrating that the motion-tracking scheme successfully follows the sample motion while acquiring the THz-TDS waveform, even though the 10-cm displacement is more than ten times larger than the sample-displacement range accessible by conventional THz-TDS without motion tracking.

For practical applications, body motion is generally not known a priori, and the respiration and heartbeat frequencies must therefore be estimated directly from the measured data. As a simple demonstration, we calculated the second time derivative of the measured displacement to enhance the oscillatory components. Before calculating the second derivative, the measured data were digitally low-pass filtered with a cutoff frequency of 5 Hz, which is well above the typical respiration and heartbeat frequencies, to suppress high-frequency feedback noise. The prescribed motions of the linear stage, used to mimic body motion, produced large acceleration spikes in the second-derivative waveform. Because these spikes appeared as isolated pulse-like features, they could be readily identified and removed. We then performed Fourier analysis of the resulting waveform. From the resulting spectrum, the respiration and heartbeat frequencies were estimated to be $f_b=0.25$ Hz and $f_h=1.2$ Hz, respectively, in good agreement with the prescribed values.
More robust extraction of respiratory and cardiac components in the presence of unknown body motion will require more sophisticated approaches, such as motion-contaminated interval rejection \cite{Khan2017}, cyclostationary analysis exploiting vital-sign periodicity \cite{Kazemi2014}, and multi-radar motion cancellation \cite{Jang2021}, and remains a subject for future work.


\section{\label{sec:level6}Conclusion}
We demonstrated motion-tracking THz-TDS based on TPFCs, enabling rapid THz waveform acquisition while continuously tracking a moving sample. Real-time apodized acquisition using TPFCs enabled the measurement of a 61-ps THz temporal window at an acquisition rate of up to 154 Hz. By utilizing both the forward and backward scans, the acquisition rate can be doubled to 308 Hz, and even higher acquisition rates can be achieved by reducing the temporal window.
The system also achieved a timing stability of 1.5 fs at an averaging time of 0.44 s, corresponding to a displacement stability of 225 nm in reflection geometry. 
Furthermore, by using one TPFC for real-time apodized waveform acquisition and the other to control the temporal origin through feedback, the measurement window was made to follow large displacements of a reflecting sample while rapidly acquiring the THz waveform. This approach enabled continuous tracking of sample motion beyond the displacement range defined by the original measurement window, while simultaneously providing displacement information from the feedback and residual peak-position signals.
As a proof-of-concept application, we performed a model demonstration of noncontact vital-sign sensing using a vibrating voice-coil motor. Respiration- and heartbeat-related oscillations were successfully identified in the presence of much larger linear-stage motion that simulated unintended body motion. 
Motion-tracking THz-TDS is expected to extend the applicability of THz-TDS to measurements in which sample motion cannot be neglected, including materials characterization and industrial nondestructive inspection.
\section*{ACKNOWLEDGMENTS}
Parts of this study are supported by JSPS KAKENHI (23K26165, 24K21743, 25KJ2061, 25H02153, 26H02144), JST CREST (Grant No. JPMJCR19J4), MEXT Q-LEAP (Grant No. JPMXS0118067246), and Precise Measurement Technology Promotion Foundation. 

\section*{AUTHOR DECLARATIONS}

\subsection*{Conflict of Interest}
The authors have no conflicts to disclose.

\subsection*{Author Contributions}
\noindent \textbf{Riku Shibata}: Conceptualization (equal), Data curation (lead), Formal analysis (lead), Investigation (lead), Methodology (lead), Software (lead), Validation (lead), Visualization (lead), Writing -- original draft (lead). \textbf{Shun Fujii}: Methodology (supporting), Validation (equal), Writing -- review \& editing (equal). \textbf{Tomofumi Ikari}: Methodology (supporting), Validation (supporting), Writing -- review \& editing (equal). \textbf{Shinichi Watanabe}: Conceptualization (lead), Data curation (equal), Formal analysis (equal), Funding acquisition (lead), Investigation (equal), Methodology (equal), Project administration (lead), Resources (lead), Supervision (lead), Validation (equal), Writing -- review \& editing (lead).

\section*{DATA AVAILABILITY}
The data that support the findings of this study are available from the corresponding author upon reasonable request.

\appendix

\section{\label{sec:appendix} Determination of Feedback Parameters Based on Step Responses}

\begin{figure}
\includegraphics[width=\linewidth]{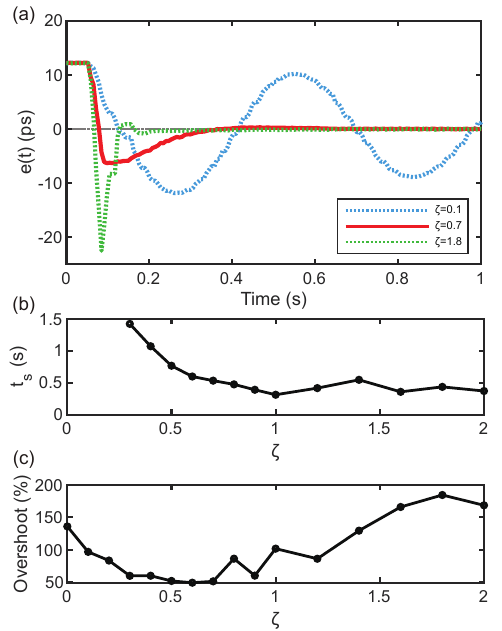}
\caption{\label{fig5} (a) Measured peak-position error $e(t)$ for representative damping ratios of $\zeta=0.1$, 0.7, and 1.8. (b) Settling time $t_s$ and (c) percent overshoot as functions of the damping ratio $\zeta$. The natural frequency was fixed at $\omega_n/2\pi=1.73$ Hz.}
\end{figure}

In this Appendix, based on the block diagram shown in Fig.~\ref{fig1}(b), we derive the error transfer function of the motion-tracking system from the time shift $d(t)$ induced by mirror displacement to the THz peak-position error $e(t)$ and experimentally determine the optimal feedback parameters. Since $d(t)$ can be intentionally introduced by moving the linear stage and $e(t)$ can be experimentally determined from the THz-TDS peak position, the transfer function provides a practical means of evaluating the feedback response and determining the feedback parameters that optimize the system performance.

First, to simplify the theoretical analysis, we approximate the feedback system as a continuous-time system and neglect the time delay unit in Process 1 in the feedback loop. From the block diagram, the continuous-time counterparts of Eqs.~\eqref{eq1}--\eqref{eq4}, with $\tau_0$ held constant, are expressed as follows:
\begin{align}
u(t)&=K_pe(t)+K_i\int_0^t e(t')dt',\label{eq:A1}\\
-e(t)&=\int_0^t u(t')dt' +d(t),\label{eq:A2}
\end{align}
where $t'$ is the integration variable. 
The Laplace transforms of $e(t)$, $u(t)$, and $d(t)$ are denoted by $E(s)$, $U(s)$, and $D(s)$, respectively, where $s$ is the Laplace variable. Then, Eqs. \eqref{eq:A1} and \eqref{eq:A2} become
\begin{align}
U(s)&=K_pE(s)+K_i \frac{E(s)}{s},\label{eq:A3}\\
-E(s)&=\frac{U(s)}{s} + D(s).\label{eq:A4}
\end{align}
Therefore, the error transfer function $G_e(s)$ from $D(s)$ to $E(s)$ is given by
\begin{align}
G_{e}(s)
=
\frac{E(s)}{D(s)}
&=
-\frac{s^2}
{s^2+K_p s+K_i} \notag\\
&=-\frac{s^2}
{s^2+2\zeta\omega_n s+\omega_n^2},
\label{eq:A5}
\end{align}
where
\begin{equation}
\omega_n=\sqrt{K_i},
\qquad
\zeta=\frac{K_p}{2\sqrt{K_i}}
\label{eq:A6}
\end{equation}
are the natural angular frequency and damping ratio of the feedback system, respectively. 

Next, we experimentally determine the optimal values of $\omega_n$ and $\zeta$ for the motion-tracking feedback system from the step response. For the step-response measurement, the THz peak position was first set at a position displaced by $\delta=-12 $~ps from the center of the time window of the forward waveform. The feedback setpoint $\tau_0$ was then set to the center of the time window, thereby introducing an initial step-like peak-position error of $\delta$, and motion-tracking THz-TDS was performed. The natural angular frequency was fixed at $\omega_n/2\pi=1.73$ Hz as a practical operating value that provided sufficiently fast tracking without causing the THz peak to leave the time window in preliminary tests. With $\omega_n$ fixed, the damping ratio $\zeta$ was varied from 0.01 to 2, and the resulting step responses were evaluated.
The accessible THz-TDS time window was set to $\pm 30.5$ ps, and the feedback interval $\Delta t$ was 6.5 ms.

Figure~\ref{fig5}(a) shows the measured peak-position error $e(t)$ for different values of $\zeta$. At $t=0$, the initial peak-position error was $e(0)=12$ ps, and the feedback system acted to drive $e(t)$ toward zero. For small $\zeta$, $e(t)$ exhibited a slowly decaying oscillatory response after crossing zero, and consequently required a relatively long time to settle. As $\zeta$ increased, this damped oscillation was suppressed and $e(t)$ converged more rapidly toward zero. At still larger $\zeta > 1 $, however, a pronounced overshoot accompanied by a rapid oscillatory transient appeared shortly after the feedback response began. 

The latter behavior with larger $\zeta$ can be attributed to the delay in the actual feedback system. As shown in Fig.~\ref{fig5}(a), $e(t)$ remained nearly unchanged until $t=t_7=7\Delta t\approx45 $~ms, independent of $\zeta$, indicating that the control action did not yet affect the measured peak position during the first eight frames, possibly due to the initial communication latency in the feedback system. Nevertheless, the control signal $u(t)$ was continuously calculated during this period. Consequently, the control commands accumulated before their effect appeared in $e(t)$, leading to an excessive correction when the system began to respond. This effect became more pronounced at larger $\zeta$, corresponding to a larger proportional gain $K_p$, and produced the rapid overshoot observed in Fig.~\ref{fig5}(a). 

To determine the optimal value of $\zeta$ for robust feedback control, we examined the settling time $t_\mathrm{s}$ and the maximum overshoot $e_{\max}$. 
The settling time $t_\mathrm{s}$ was defined as the time at which $|e(t)|$ first decreased below 2\% of the initial error, corresponding to 0.24 ps, and subsequently remained within this range. 
Figure~\ref{fig5}(b) shows the dependence of $t_{\mathrm{s}}$ on $\zeta$ for the cases in which settling was observed. The settling time decreased with increasing $\zeta$ and reached its minimum at $\zeta=1$.
The maximum overshoot $e_{\max}$ was defined as the maximum value of each measured $|e(t)|$ and used as a measure of the system stability. Then the percent overshoot $M_p$ is calculated as, 
\begin{equation}
M_p (\%) = \left|\frac{e_\mathrm{max}}{\delta}\right|\times100.
\label{eq:A7}
\end{equation}
Figure~\ref{fig5}(c) shows the dependence of $M_p$ on $\zeta$. The overshoot was large at both small and large $\zeta$, with a minimum at $\zeta=0.6$. 
We selected $\zeta=0.7$ for this study, as it provided an $M_p$ close to the minimum while maintaining a short $t_\mathrm{s}$.

\section{\label{sec:appendix}Frequency response of the Motion-Tracking Control System.}
\begin{figure*}
\includegraphics[width=\linewidth]{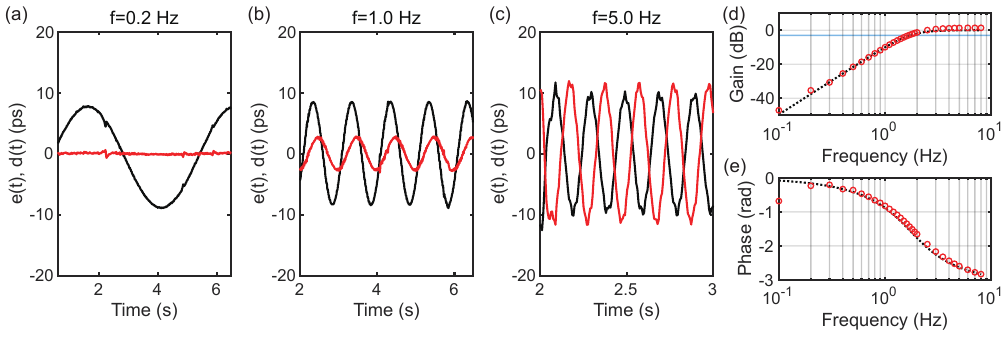}
\caption{\label{fig6} (a)–(c) Measured peak-position error $e(t)$ (red curves) and reconstructed temporal shift $d(t)$ (black curves) for VCM oscillation frequencies of 0.2, 1.0, and 5.0 Hz, respectively. (d) Error gain and (e) phase difference between $e(t)$ and $d(t)$ as functions of the VCM oscillation frequency. Red circles and black dashed curves represent the experimental results and theoretical predictions, respectively. The blue horizontal line in (d) indicates the -3 dB level.}
\end{figure*}

In this Appendix, we evaluated the frequency response of the motion-tracking control system and estimated the frequency range over which motion-tracking is effective. The experimental setup was the same as that described in Sec. III-C. Here, the VCM was driven sinusoidally at an oscillation frequency $f$ varied from 0.1 to 8~Hz, with a constant amplitude of 1.5 mm, corresponding to approximately 10 ps in the THz waveform. The linear stage was fixed at a constant position during the measurement. The other experimental parameters were set to the optimal values determined in Appendix A.

Figures~\ref{fig6}(a)–(c) show representative measurement results for VCM oscillation frequencies $f$ of 0.2, 1.0, and 5.0 Hz, respectively. In each panel, the red and black curves represent measured $e(t)$ and reconstructed $d(t)$, respectively. 
At VCM frequencies of 0.2 Hz (Fig.~\ref{fig6}(a)), the control signal effectively compensated for the tracking error $e(t)$, keeping it close to zero after the initial transient and thereby achieving good tracking performance.
As the VCM frequency increased, an oscillatory component gradually emerged in $e(t)$, accompanied by an increasing phase lag.
At 5.0 Hz in Fig.~\ref{fig6}(c), the amplitude of $e(t)$ became comparable to that of $d(t)$, and the phase lag approached $-\pi$, resulting in nearly antiphase behavior. At this frequency, the control signal $u(t)$ approached zero (see Eq. (\ref{eq5})), indicating that the motion-tracking control was no longer effective. 

To investigate the frequency response from $d(t)$ to $e(t)$ in detail, we constructed Bode plots of the corresponding magnitude and phase responses as follows. At each frequency $f$, the steady-state portions of the measured $e(t)$ and $d(t)$, excluding the initial transient, were fitted using
\begin{equation}
A_i\sin(2\pi ft+\phi_i)+C_i,\quad i=e,d,
\label{eq:B1}
\end{equation}
where $i=e,d$ denotes $e(t)$ and $d(t)$, respectively, and $A_i$, $\phi_i$, and $C_i$ are the corresponding amplitude, phase, and DC offset. Using the obtained amplitudes and phases, we defined the experimental error gain $L(f)$ and phase $\Delta \phi(f)$ as, 
\begin{gather} L(f)=20\log_{10}\left(\frac{A_e}{A_d}\right), \label{eq:B2} \\ \Delta\phi(f)=\phi_e-\phi_d, \label{eq:B3} 
\end{gather}
respectively. 
The red circles in Figs.~\ref{fig6}(d) shows the resulting error gain $L(f)$. At low frequencies, the gain $L(f)$ is small, whereas it approaches 0 dB as the frequency increases, providing a frequency-domain representation of the behavior discussed above in Figs.~\ref{fig6}(a) --(c).

For comparison, the theoretical frequency response was obtained by substituting $s=i\omega$, with $\omega=2\pi f$, into the error transfer function in Eq.~\eqref{eq:A5}:
\begin{equation}
G_e(i\omega)=-\frac{(i\omega)^2}{(i\omega)^2+2\zeta\omega_n(i\omega)+\omega_n^2}.
\label{eq:B4}
\end{equation}
The corresponding error gain is then calculated as $20\log_{10}|G_e(i\omega)|$. The black dashed curve in Fig.~\ref{fig6}(d) shows the theoretical error gain, which agrees well with the experimental results. We defined the cutoff frequency $f_c$ as the frequency at which the error gain reaches $-3$ dB. The cutoff frequency was $f_c=1.71$ Hz, below which the motion-tracking control attenuates $e(t)$ in response to $d(t)$.

The red circles in Fig.~\ref{fig6}(e) show the experimentally obtained phase $\Delta\phi(f)$. At low frequencies, the phase was close to 0, i.e., the $e(t)$ is approximately in phase with $d(t)$, consistent with the effective motion tracking indicated by the small error gain. At high frequencies, the phase approaches $-\pi$, indicating that $e(t)$ becomes approximately opposite in phase to $d(t)$, consistent with the loss of effective motion tracking. As with the error gain, we derived the theoretical phase response from the transfer function in Eq.~\eqref{eq:B4} by calculating $\arg G_e(i\omega)$. The black dashed line in Fig.~\ref{fig6}(e) shows the theoretical phase response, which agrees well with the experimental results. 

\section{\label{sec:appendix}Theoretical Analysis of the Ramp Response of the Motion-Tracking Control System}

In this Appendix, we theoretically analyze the transient response of $e(t)$ observed in Section III.B when the mirror undergoes ramp-like motion.
The temporal shift induced by the mirror motion, as described in Fig. 3(b), can be expressed by the following ramp input:
\begin{equation}
d(t)=
\begin{cases}
0,
& t<t_{\mathrm{b}},\\[4pt]
v_d(t-t_{\mathrm{b}}),
& t_{\mathrm{b}}\leq t<t_{\mathrm{e}},\\[4pt]
v_d(t_{\mathrm{e}}-t_{\mathrm{b}}),
& t\geq t_{\mathrm{e}},
\end{cases}
\label{eq:C1}
\end{equation}
where $t_{\mathrm{b}}$ and $t_{\mathrm{e}}$ denote the times at which the mirror starts and stops moving, respectively. Here, $v_d$ is the rate of change of the THz delay time. For a mirror moving along the $z$-axis at a velocity $v$, it is given by $v_d=2v/c$.
The peak-position error $e(t)$ is derived from following inverse Laplace transform:
$
e(t)=
\mathcal{L}^{-1}\{G_{e}(s)D(s)\}
\label{eq:C2}.
$
For an underdamped system with $0<\zeta<1$, $e(t)$ can be expressed as
\begin{widetext}
\begin{equation}
e(t)=
\begin{cases}
0,
& t<t_{\mathrm{b}},\\[6pt]
-\dfrac{v_d}{\sqrt{1-\zeta^2}\,\omega_n}
\exp\left[-\zeta\omega_n(t-t_{\mathrm{b}})\right]
\sin\left[
\sqrt{1-\zeta^2}\,\omega_n(t-t_{\mathrm{b}})
\right],
& t_{\mathrm{b}}\leq t<t_{\mathrm{e}},\\[10pt]
\dfrac{v_d}{\sqrt{1-\zeta^2}\,\omega_n}
\exp\left[-\zeta\omega_n(t-t_{\mathrm{e}})\right]
\sin\left[
\sqrt{1-\zeta^2}\,\omega_n(t-t_{\mathrm{e}})
\right],
& t\geq t_{\mathrm{e}}.
\end{cases}
\label{eq:C2}
\end{equation}
\end{widetext}
Here, the transient generated at $t=t_{\mathrm{b}}$ is assumed to have sufficiently decayed by $t=t_{\mathrm{e}}$, i.e., $t_e-t_b\gtrsim4/\zeta\omega_n$, where $4/\zeta\omega_n$ is the 2\% settling-time. 
The black dashed curve in Fig.~\ref{fig3}(a) shows $e(t)$ calculated under the experimental conditions described in Sec.~III-B.
The waveform of $e(t)$ exhibits two peaks near $t=t_s=2$~s and $t=t_e=5$~s and closely reproduces the experimentally observed error signal. The slight discrepancies can be attributed to the time delay associated with Process~1, which was neglected in the above theoretical analysis, and to the approximation of $d(t)$ that neglects the stage acceleration.

The absolute value of the peak of $e(t)$ is given by,
\begin{align}
\left|e\right|_{\max}
&=
\frac{v_d}{\omega_n}
\exp\left[
-\frac{\zeta}{\sqrt{1-\zeta^2}}
\tan^{-1}\left(
\frac{\sqrt{1-\zeta^2}}{\zeta}
\right)
\right].
\label{eq:C3}
\end{align}
In motion-tracking THz-TDS, if $e_{\max}$ exceeds half of the time window, the THz peak may fall outside the measurable time window when the sample velocity changes abruptly, preventing continuous measurement. Therefore, it is desirable to minimize $e_{\max}$. For the experimental conditions used in this study, namely a temporal window of $\pm 30.5$ ps, $\omega_n/2\pi=1.73$ Hz, and $\zeta=0.7$, Eq.~\eqref{eq:C3} gives a maximum trackable delay velocity of approximately $v_d=723$ ps/s, corresponding to a sample velocity of approximately 10.8 cm/s in reflection geometry. As indicated by Eq.~\eqref{eq:C3}, the peak error decreases with increasing $\omega_n$ and $\zeta$, appropriate to minimize $e_{\max}$. 
However, in the actual feedback system, the time delay in Process~1 and discrete-time effects, which are neglected in the above analysis, introduce additional phase lag that becomes increasingly significant as $\omega_n$ and $\zeta$ increase, thereby increasing $e_{\max}$ and reducing the stability margin.
Therefore, $\omega_n$ and $\zeta$ must be selected by taking into account the stability of the actual feedback system.

\section*{References}\bibliography{aipsamp}

\end{document}